\documentclass[9pt,a4paper,notitlepage,twocolumn,oneside]{article}

\usepackage{graphicx}

\title{Particle-stabilized oscillating diver: a self-assembled responsive capsule}
\author{Joseph W.~Tavacoli, Job H.~J.~Thijssen and Paul S.~Clegg$^*$}

\begin{document}

\maketitle

\noindent{\footnotesize{\textit{School of Physics and Astronomy, University of Edinburgh, Mayfield Road, Edinburgh EH9 3JZ, United Kingdom. E-mail: paul.clegg@ed.ac.uk}}}

\noindent \textbf{We report the experimental discovery of a self-assembled capsule, with density set by interfacial glass beads and an internal bubble, that automatically performs regular oscillations up and down a vial in response to a temperature gradient. Similar composites featuring interfacial particles and multiple internal compartments could be the solution to a variety of application challenges.}

\vspace{0.5cm}

Recent experiments have shown that well-designed Pickering emulsions and foams (in which the droplets or bubbles are stabilized by interfacially trapped solid particles~\cite{Binks06}) can be triggered to move~\cite{Melle05, Brugger07, Zhao09} and/or disintegrate~\cite{Read04, Ngai06, Saigal10, Hwang10} in response to an external stimulus (pH response for liquid marbles~\cite{Dupin2009} and polymer-stabilized emulsions~\cite{Weaver2009} has also been shown). Furthermore microfluidic techniques~\cite{Chu07, Wan08}, multi-stage mixing~\cite{Aveyard03} and liquid-liquid phase separation~\cite{Clegg07} have all been harnessed to prepare emulsions with multiple internal compartments with well controlled sizes and compositions. Combining interfacial particles that couple to an external stimulus, while also incorporating multiple internal compartments, holds the promise of revolutionizing the capabilities of emulsions. Separate fluid and solid components can provide sophisticated responses (e.g. to temperature, solvent environment, etc.) with well-controlled time signatures. Here we report the combination of interfacial particles, that tune the droplet's response to gravity, with two internal compartments (one liquid, one gas) that create sensitivity to temperature. We demonstrate that the droplet exhibits a complex response to an externally imposed temperature gradient (periodic diving). Intriguingly, our composite droplet self assembles.

Formation begins with a vial, Fig. 1(I), holding a lower liquid (ethanediol) and an overlying layer of a less dense, more volatile liquid (pentane). The particles (glass beads, diameter $d \approx 400$\,$\mu$m) are then poured in, with the majority falling through the liquid interfaces to the base of the vial. As they do so they entrain both pentane and air, out of which the capsule forms, Fig. 1(I). Initially the pentane droplet with a gas bubble inside is smothered by overlying particles making the capsule negatively buoyant (see Fig. 2 and Supplementary Information including Movie 1). However, further expansion of the bubble and loss of excess particles inverts the buoyancy and the capsule begins to rise. Particles trapped at the capsule interface augment the mass of the droplet while the growing bubble reduces it. As shown in Fig. 1(II), the capsule begins to move, first rising and then sinking down the vial; the rising and sinking is repeated with a regular period for many cycles (see also Supplementary Information, Movie 2). We call this automatic diving capsule that has two internal compartments (one liquid, one gas, Fig.~1(III)) the particle-stabilized oscillating diver (POD).

The regular oscillations of the POD continue until the pentane overlayer has completely evaporated (see Supplementary Information, Movie 3). This process slowly cools the top of the vial, generating a temperature gradient. In typical experiments, carried out at room temperature, the top of the continuous phase is 3K cooler than the base. When the pentane is exhausted the pattern of sinking and rising can be re-instigated by other means; e.g. a diver that has come to rest can begin moving again when a piece of ice is floated at the surface. The POD motion is driven by its environment (i.e. the temperature gradient) via the bubble shrinking at the top of the vial and swelling at the bottom (Fig. 3a, b). The change in volume reflects, respectively, the condensation and evaporation of pentane. (The expansion of the air in the bubble in response to the change in temperature makes a small contribution to the increase in bubble size.) These volume changes take place over several seconds, and hence lag behind the change in height, Fig. 3c, leading to a situation where the POD never reaches thermal equilibrium with the background phase. It is the time lag that perpetuates the oscillation. 

The POD is conceptually related to the well known Cartesian diver. Originally discovered by a student of Galileo, the Cartesian diver is a capsule containing an air bubble that floats inside an enclosed volume of water. The diver sinks when the water volume is squeezed as a consequence of the compression of the trapped air. It was pointed out half-a-century ago that a Cartesian diver can be made that will perpetually travel up and down a vial without intervention~\cite{Mackay58a, Mackay58b}. The diver must contain a volatile solvent and a temperature gradient must be maintained between the top (colder) and the bottom (warmer) of the vial. The device is called an automatic Cartesian diver and our POD is the first self-assembled version, Fig. 1 and 2. More broadly, the pattern of behavior we have described is characteristic of a class of dynamical systems known as relaxation oscillators~\cite{Grasman87}. 

Next we explain the size and the oscillation period of our POD. We then show how these change when we choose different constituents (see also Supplementary Information). The thin shell of particles controls the size (and alters the average density) of the capsule. Initially, during formation, the particles are trapped at the flat interface between the two liquids until this interface is destabilized by the additional weight (Fig. 1(II) b, c). The weight of the region that falls has to be large enough to overcome the restoring force imposed by the interfacial tension between the two liquids. An estimate is given by the capillary length, $\lambda$, for an interface between two liquids with an interfacial tension corresponding to that between pentane and ethanediol ($\eta = 15.5 \pm 0.2$\,mN m$^{-1}$) and a density difference similar to that between the particles and ethanediol ($\Delta \rho \approx 1400$\,Kg m$^{-3}$). This gives $\lambda \approx 1$\,mm; i.e. about one third of the typical POD size reflecting the fact that we have not attempted to estimate the mass distribution of the combined pentane and glass beads. 

For many applications control of the speed of response is essential. For the POD, changes in the solvent temperature give rise to height changes and these in turn switch the POD from heating to cooling. The complete cycle has a period $\tau_{exp} = 43 \pm 5$\,sec (e.g. Fig. 3c). It will vary with the product of the heat capacity and the thermal resistance of the solvent as this is the time needed for the POD temperature to change (the depth of the background liquid, which we keep constant, also plays a role). We estimate, Fig. 3, that the complete oscillation period should be $\tau_c = 2 t_J + \tau \approx 57$\,sec in reasonable agreement with $\tau_{exp}$. The time spent switching between `up' and `down' states can be tuned by altering the viscosity of the surrounding solvent.

To test the versatility and robustness of the POD concept, and to demonstrate its potential in environmentally more benign systems we have repeated our POD fabrication with the continuous phase, ethanediol, replaced with water; smaller particles are used together with additional surfactant to adjust the three-phase contact angle (see Supplementary Information). The modifications to the system (density difference, interfacial tension, particle size and viscosity) show how the behavior can be tuned: the water based POD is presented in Supplementary Information, Figure S3. The viscosity of the continuous phase is now one order of magnitude lower leading to a significant decrease in the oscillation period due to the decrease in the switching speed; there is an additional contribution due to a lowering of the heat capacity of the water based PODs (which is a result of using smaller glass beads).

We now consider further aspects of the POD as a dynamical system. Figure 2 shows the detailed processes by which the mixture of liquid, gas and particles at the base of the vial becomes buoyant. During formation, the rising liquid and gas push some particles aside; but it is those that become trapped on the liquid-liquid interface that are of most interest (Fig. 2d). Here the large glass beads form a granular medium that is trapped on a two dimensional surface. All of the beads are forced down by gravity (Fig. 2d, a small number of beads are colored to ease visualization); the collective weight of higher beads on lower ones appears sufficient to eject lower ones individually from the interface in spite of the strength of the trapping (see also Supplementary Information, Movie 1).

A further surprise is that, as well as displaying oscillatory motion in the vertical plane, the POD also rotates about its vertical axis. The rotation takes place at the bottom of the cell, simultaneously with the bubble expansion, and ceases as soon as the POD begins to rise (see Supplementary Information, Movie 4). In all cases of rotation, the POD appears to be in contact with the particle bed. This may not be significant if rotation is a function of depth as appears possible: on shallow dives no rotation is witnessed. Rotation in both directions is seen but an individual POD often rotates only in one direction. Currently we cannot explain this behavior but one conjecture is that the POD is caught up in convection currents at the bottom of the vial.

Our understanding of the POD as a self-assembled automatic Cartesian diver can guide our search for new functionality. We now know that by changing parameter values different regimes of behavior can be harnessed. For example, a bistable capsule would be better aligned to triggered delivery applications than one that oscillates. A larger bubble, a larger temperature gradient or a more dramatic thermal expansion could be chosen to give a capsule that would rise and stick to the top of a sample as the top cools (we have observed the larger bubble case on several occasions). Furthermore, a coupling between temperature and density is only one of numerous possible pairings. For example, a composite droplet where changes in pressure at different heights in a vessel couple appropriately to the volume of an internal compartment could be used to address emulsion creaming / sedimentation problems. 

In summary, we have observed the formation of a composite capsule that sinks and rises in response to a temperature gradient. This diving capsule self assembles when particles are poured into a vial containing two immiscible liquids. A thin shell of particles is essential to tuning the average density of the capsule while the volatile solvent, trapped bubble and temperature gradient create height dependent buoyancy. From these observations we conclude, more generally, that particle-stabilized droplets can be fabricated with complex responses to external conditions.

We are grateful to M. Cates, S. Egelhaaf and W. Poon for helpful comments and to EPSRC GR/E030173/01, SE/POC/8-CHM-002 and the COMPLOIDs EU Initial Training Network for funding.

\begin{figure*}
\centering
\includegraphics[width=0.95\textwidth]{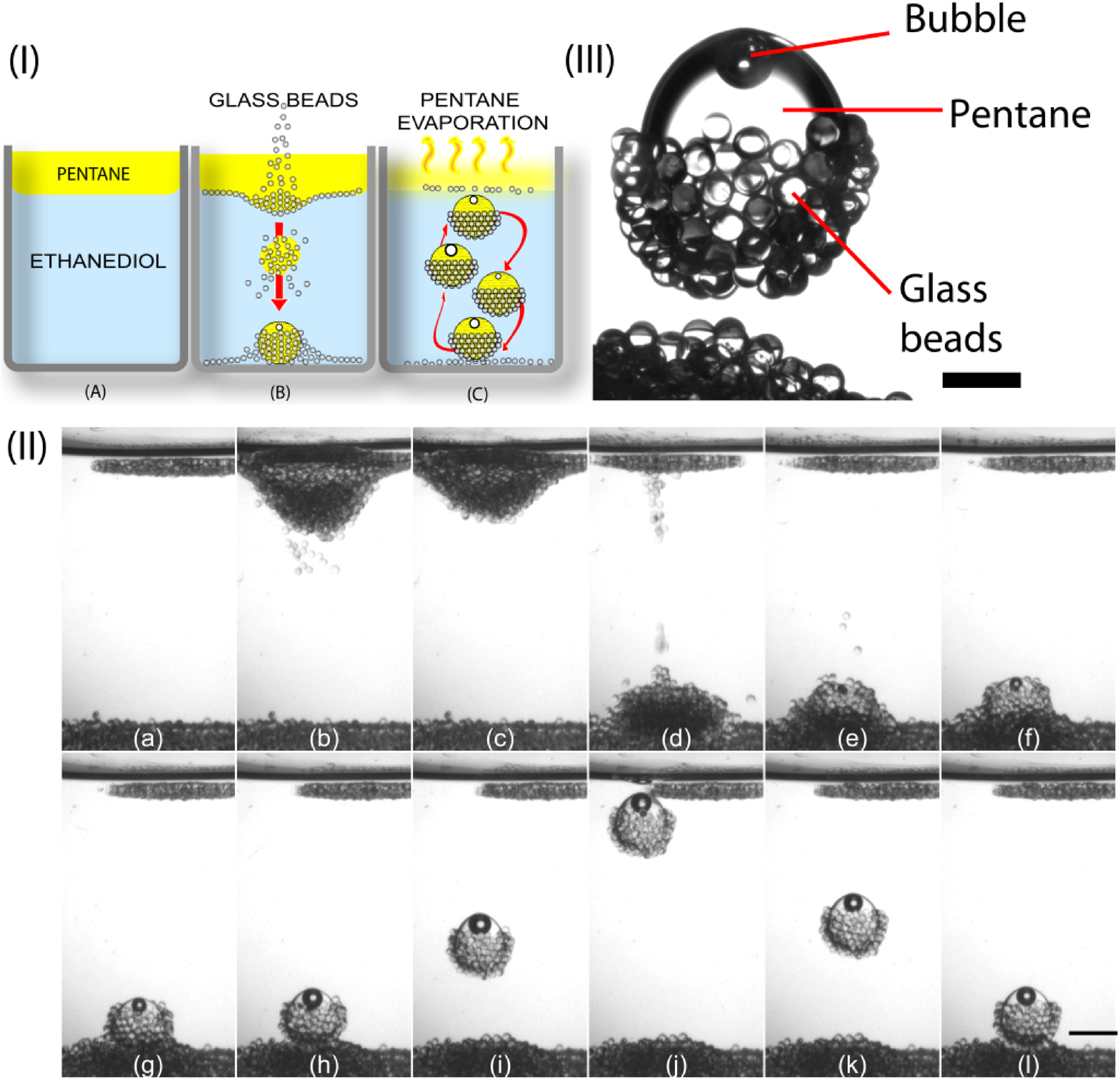}
\caption{Preparation of a particle-stabilized oscillating diver. (I) Two immiscible solvents pentane and ethanediol are put in a vial (A). It is important that the top one is volatile. Subsequently the beads are poured in using a spatula (B). Oscillations then emerge (C). (II: a-l) Frames from a Movie showing formation taking place, scale bar 3 mm. (II: a-e) Show the process of self-assembly; (II: f-l) show the first complete oscillation. The time of frames (b) to (l), with respect to frame (a), are 2.0, 2.5, 3.0, 5.0, 7.5, 10.0, 30.5, 48.5, 61.5, 86.5, 91.0 and 107.0 s. (III) Shows the two internal compartments (pentane and bubble) together with the layer of trapped beads. Scale bar 1mm. }
\label{fig:BirthofthePOD}
\end{figure*}

\begin{figure*}
\centering
\includegraphics[width=0.95\textwidth]{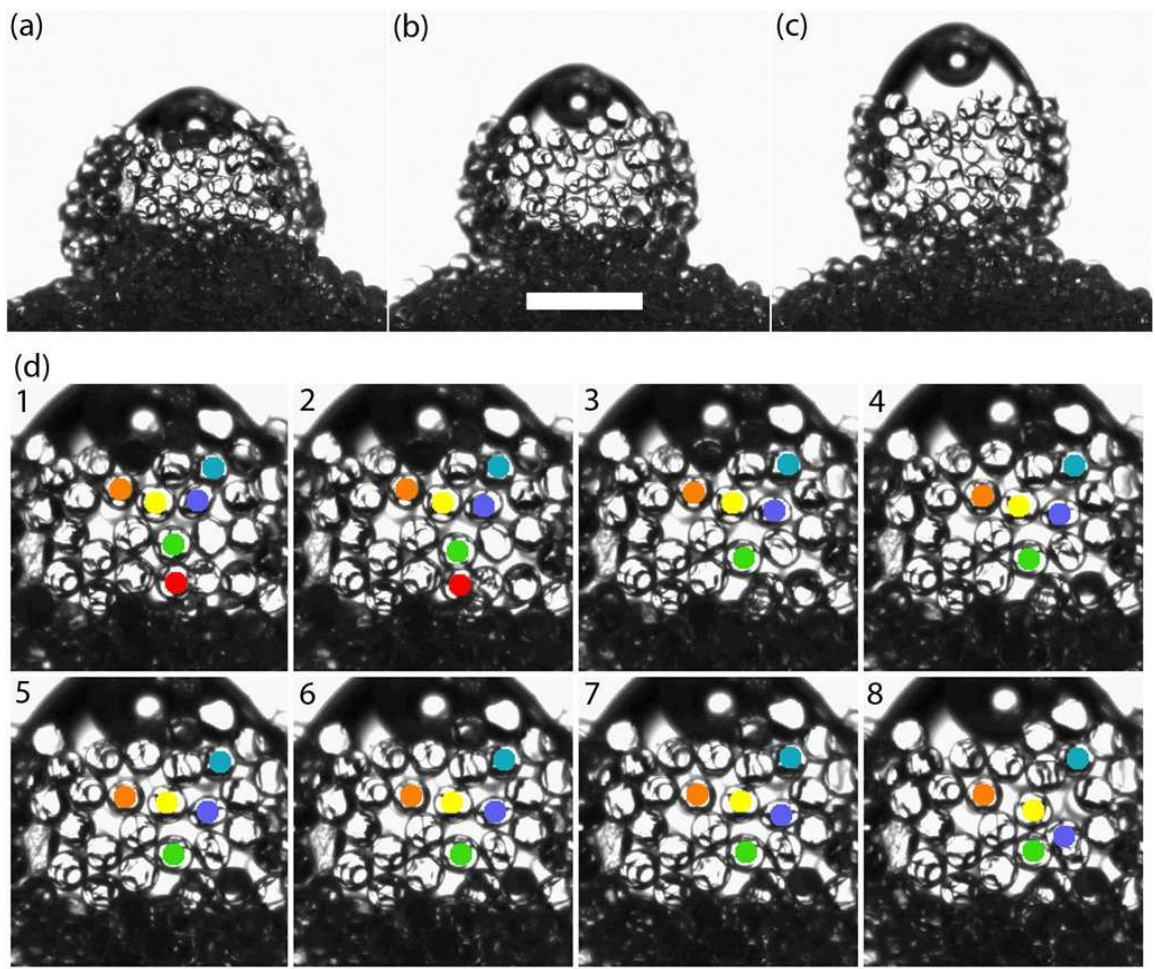}
\caption{The emergence of a buoyant POD at the base of the vial. (a-c) Frames from a movie (see Supplementary Information, Movie 1) showing the pentane droplet and bubble pushing upward through the pile of beads. As the POD begins to form glass beads are pushed aside and a small number are ejected from the liquid-liquid interface. Scale bar 2 mm; images separated by 6 sec. (d) Zoomed images every 0.25 sec from a segment between the frames shown in (a) and (b) with selected beads in color. The red bead is ejected between images 2 and 3; the purple bead has detached and is falling in image 8. As a result of these events the relative positions of the remaining beads has altered significantly. Through a progressive series of rearrangements the POD becomes buoyant.}
\label{fig:Emergence}
\end{figure*}

\begin{figure*}
\centering
\includegraphics[width=0.95\textwidth]{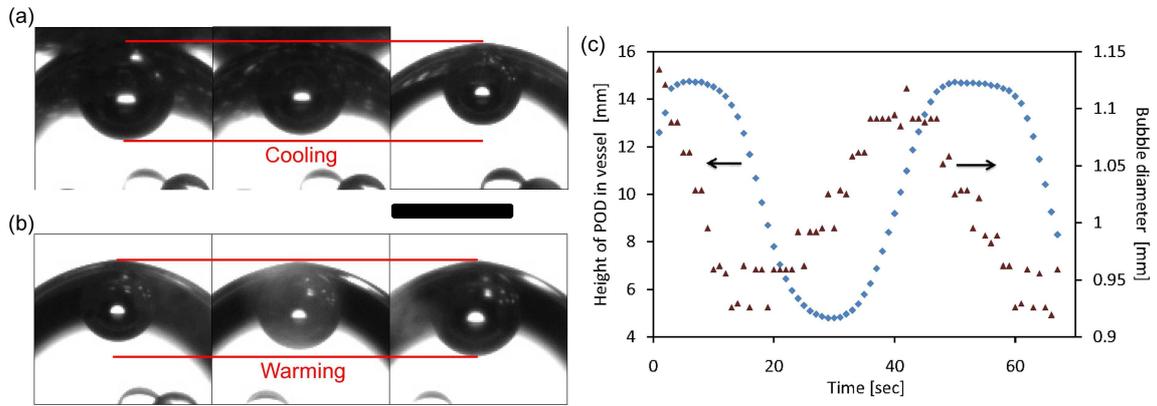}
\caption{POD oscillation mechanism and its time structure. (a) Successive frames (left to right) showing the bubble deflating at the cooler top of the vial and, (b), inflating at the warmer base. The scale bar is 1mm. (c) Height of POD in vial (blue diamonds) and corresponding bubble diameter (burgundy triangles) as a function of time. This demonstrates the substantial time lag between the change of bubble size and the change of POD location. We estimate that the heat capacity of the POD is 0.02\,J\,K$^{-1}$ and the thermal resistance is 1350\,KW$^{-1}$ (based on a thermal conductivity of pentane $k = 0.12 $W\,m$^{-1}$\,K$^{-1}$). The product suggests a period $\tau = 27$\,sec  which does not yet include the time during which the POD is moving up and down. Averaged over 14 oscillations we find that the velocity of rising is $v_{rise} = 0.9 \pm 0.1$\,mm\,s$^{-1}$ and of falling is $v_{fall} = 0.9 \pm 0.3$\,mm\,s$^{-1}$ (the first oscillation can be anomalous). For the height of ethanediol used, the journey from, say, top to bottom takes the POD about $t_J \approx 15$\,sec. The complete oscillation period should be $\tau_c = 2t_J + \tau \approx 57$\,sec in reasonable agreement with observations (c).}
\label{fig:Oscillations}
\end{figure*}

\end{document}